\documentclass[twocolumn,reprint,aps,prl,amsmath,amssymb]{revtex4-1}
\usepackage[latin9]{inputenc}
\usepackage{amsmath}
\usepackage{graphicx}
\usepackage{esint}

\makeatletter

\@ifundefined{textcolor}{}
{%
 \definecolor{BLACK}{gray}{0}
 \definecolor{WHITE}{gray}{1}
 \definecolor{RED}{rgb}{1,0,0}
 \definecolor{GREEN}{rgb}{0,1,0}
 \definecolor{BLUE}{rgb}{0,0,1}
 \definecolor{CYAN}{cmyk}{1,0,0,0}
 \definecolor{MAGENTA}{cmyk}{0,1,0,0}
 \definecolor{YELLOW}{cmyk}{0,0,1,0}
}

\usepackage{braket}

\renewcommand{\vec}{\mathbf}                     
\newcommand{\pd}[2]{\frac{\partial #1}{\partial #2}}
\newcommand{\pdn}[3]{\frac{\partial^{#3} #1}{{\partial #2}^{#3}}}
\renewcommand{\Im}{\mathop{\mathrm{Im}}}

\renewcommand{\log}{{\,\mathrm{ln}}}                      

\makeatother

\begin{document}

\title{Antagonistic in-plane resistivity anisotropies from competing fluctuations
in underdoped cuprates}

\author{Michael Schütt}

\affiliation{School of Physics and Astronomy, University of Minnesota, Minneapolis
55455, USA}

\author{Rafael M. Fernandes}

\affiliation{School of Physics and Astronomy, University of Minnesota, Minneapolis
55455, USA}
\begin{abstract}
One of the prime manifestations of an anisotropic electronic state
in underdoped cuprates is the in-plane resistivity anisotropy $\Delta\rho\equiv\left(\rho_{a}-\rho_{b}\right)/\rho_{b}$.
Here we use a Boltzmann-equation approach to compute the contribution
to $\Delta\rho$ arising from scattering by anisotropic charge and
spin fluctuations, which have been recently observed experimentally.
While the anisotropy in the charge fluctuations is manifested in the
correlation length, the anisotropy in the spin fluctuations emerges
only in the structure factor. As a result, we find that spin fluctuations
favor $\Delta\rho>0$, whereas charge fluctuations promote $\Delta\rho<0$,
which are both consistent with the doping dependence of $\Delta\rho$
observed in YBa$_{2}$Cu$_{3}$O$_{7}$. We also discuss the role
played by CuO chains in these materials, and propose transport experiments
in strained HgBa$_{2}$CuO$_{4}$ and Nd$_{2}$CuO$_{4}$ to probe
directly the different resistivity anisotropy regimes. 
\end{abstract}
\maketitle
The existence of a sizable in-plane electronic anisotropy in different
families of underdoped cuprates has been established by a variety
of experimental probes, such as transport measurements \cite{Ando02,Taillefer10,Goldman13},
x-ray \cite{Hayden_PRB14,LeTacon14} and neutron scattering \cite{Hinkov08,Keimer_NJP_10},
and scanning tunneling microscopy \cite{Lawler10}. Consonant with
the proposal of electronic nematic order \cite{Kivelson98,KivelsonRMP,Vojta09,Fradkin_review},
in which the point group symmetry of the system is lowered spontaneously
by electronic degrees of freedom, these experiments provide invaluable
information for the hotly debated topic of whether any symmetries
are broken in the pseudogap phase \cite{Greven08,Shekhter13}. To
elucidate the relevance of these anisotropic properties to the phase
diagram of the cuprates, it is fundamental to establish their microscopic
origin. In this regard, a useful benchmark for theoretical proposals
is the in-plane resistivity anisotropy $\Delta\rho\equiv\left(\rho_{a}-\rho_{b}\right)/\rho_{b}$,
which was measured in the seminal work \cite{Ando02} across the phase
diagram of YBa$_{2}$Cu$_{3}$O$_{7}$ (YBCO). The moderate values
of the resistivity anisotropy that were observed experimentally, $\Delta\rho\lesssim1.5$,
are difficult to reconcile with a scenario in which metallic static
stripes \cite{Zaanen89,Machida89} order in an insulating background.
Instead, they seem to be more compatible with fluctuations that break
the tetragonal symmetry of the system \cite{KivelsonRMP,Fernandes08}.

\begin{figure}
\includegraphics[width=0.9\columnwidth]{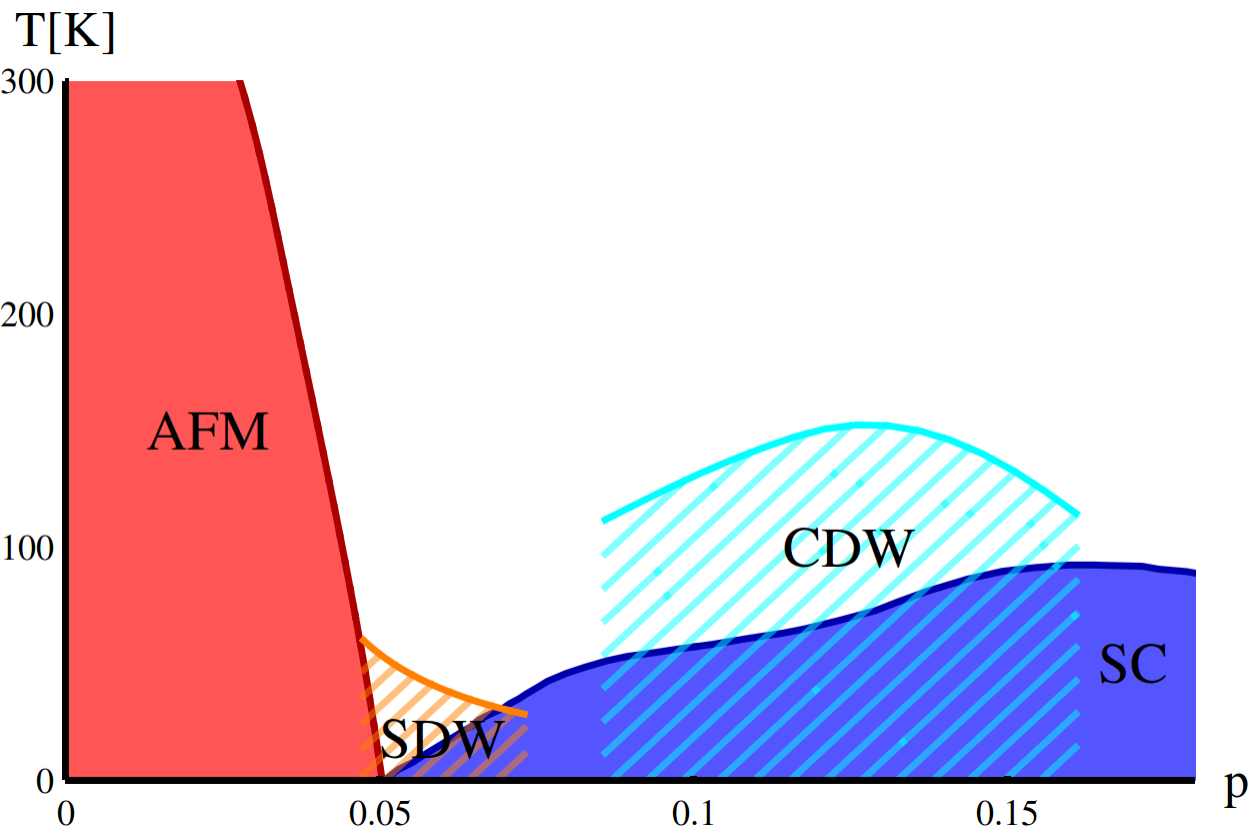} \protect\protect\protect\protect\protect\caption{Schematic phase diagram of the underdoped cuprates. Long-range incommensurate
metallic spin-density wave (SDW) order sets in at low temperatures,
next to the Mott insulating anti-ferromagnetic (AFM) phase, but its
anisotropic fluctuations persist to higher temperatures. Charge-density
wave (CDW) fluctuations, with no long-range order, are observed near
the $p=0.125$ concentration, where superconductivity (SC) is suppressed.}

\label{FigPhaseDiag} 
\end{figure}

Interestingly, neutron and x-ray measurements in underdoped YBCO have
unveiled the onset of anisotropic charge and spin fluctuations at
temperatures comparable to those marking the onset of $\Delta\rho$.
Refs.~\cite{Hinkov08,Keimer_NJP_10} found that the dynamic spin
susceptibility $\chi_{\mathrm{S}}\left(\mathbf{q},\omega\right)$
in the vicinity of the magnetic ordering vector $\mathbf{Q}_{S}=\left(\pi,\pi\right)$
becomes strongly anisotropic as temperature is lowered, eventually
giving rise to incommensurate peaks along the $a$ direction only,
and to long-range spin-density wave (SDW) order at low temperatures.
More recently, it was reported that the charge susceptibility $\chi_{\mathrm{C}}\left(\mathbf{q},\omega\right)$
is also anisotropic, with fluctuations peaked at the ordering vector
$\mathbf{Q}_{C,b}=Q_{C}\hat{\mathbf{b}}$ stronger than the fluctuations
peaked at the $90^{\circ}$-rotated ordering vector $\mathbf{Q}_{C,a}=Q_{C}\hat{\mathbf{a}}$
\cite{Hayden_PRL13,Keimer_PRL13,Hayden_PRB14,LeTacon14}. At high
magnetic fields, superconductivity is destroyed and these fluctuations
are believed to give rise to charge-density wave (CDW) order \cite{Julien11,Wu2015}.
Interestingly, the SDW and CDW fluctuations seem anti-correlated in
the phase diagram of YBCO \cite{Keimer_PRL13,Hayden_PRB14} (see Fig.
\ref{FigPhaseDiag}): while the anisotropic spin fluctuations dominate
the hole-doping concentration range $0.05\lesssim p\lesssim0.08$,
the anisotropic charge fluctuations are observed predominantly in
the $0.09\lesssim p\lesssim0.13$ range.

In this paper, we calculate the resistivity anisotropy due to the
scattering by the anisotropic charge and spin fluctuations observed
in Refs.~\cite{Hinkov08,Hayden_PRB14,LeTacon14} and compare it qualitatively
with the resistivity anisotropy measurements of Ref.~\cite{Ando02}.
Because our focus is on the sign of $\Delta\rho\equiv\left(\rho_{a}-\rho_{b}\right)/\rho_{b}$
and on its dependence on the charge and spin correlation lengths $\xi_{C}$
and $\xi_{S}$, respectively, we employ a Boltzmann equation approach.
We find that while scattering by charge fluctuations yields $\Delta\rho<0$
and $\left|\Delta\rho\right|\propto\xi_{C}^{2}$ , scattering by spin
fluctuations gives $\Delta\rho>0$ and $\left|\Delta\rho\right|\propto\ln\xi_{S}$.
These different behaviors arise from the fact that the former is governed
by the Fermi velocity at the CDW hot spots, whereas the latter is
sensitive to the curvature of the Fermi surface near the SDW hot spots.
We discuss the key role played by the CuO chains present in YBCO,
which act effectively as a conjugate field to the nematic order parameter,
selecting the experimentally-observed fluctuation anisotropies. Our
findings are consistent with the resistivity anisotropy measurements
in YBCO, and in particular with the doping dependence of $\Delta\rho$
in the range $0.05\lesssim p\lesssim0.15$.

\begin{figure}
\includegraphics[width=0.9\columnwidth]{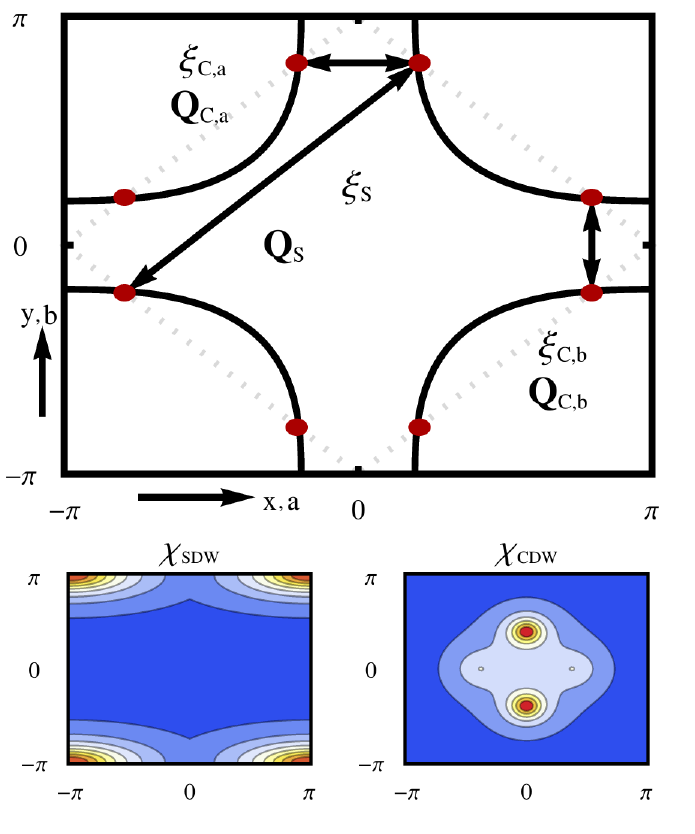} \protect\protect\protect\protect\protect\caption{(upper panel) Schematic representation of the scattering by charge
and spin fluctuations. The red dots are the magnetic hot spots. Here,
$\mathbf{Q}_{C,a(b)}=Q_{C}\hat{\mathbf{a}}(\hat{\mathbf{b}})$ and
$\mathbf{Q}_{S}=\left(\pi,\pi\right)$ correspond to the CDW/SDW ordering
vectors, and $\xi_{C,S}$ to the CDW/SDW correlation lengths. (lower
panel) Contour plots of the CDW and SDW susceptibilities given by
Eq. (\ref{EqAniSuscept}) across the first Brillouin zone, with $\eta_{S}<0$
and $\eta_{C}>0$, in accordance to experiments in YBCO.}

\label{FigScatteringMechanisms} 
\end{figure}

Our focus here is not on the mechanism responsible for the anisotropic
CDW and SDW fluctuations -- in fact, several models for nematicity
in the cuprates have been proposed \cite{Kivelson98,Oganesyan01,Kivelson_Fradkin_Geballe,JP_Kivelson08,Sun10,Fischer_Kim11,Kivelson14_disorder,Kampf13,Chubukov14,Fradkin14}.
Instead, we assume spontaneous nematic order and adopt a phenomenological
approach in which the low-energy properties of the CDW and SDW susceptibilities
are extracted from the scattering experiments \cite{Hinkov08,Hayden_PRB14,LeTacon14}.
Following previous works \cite{Sachdev_Metlitski,Efetov13,Sachdev_LaPlaca,Sachdev_Allais14,Chubukov14},
we consider the CDW ordering vectors $\mathbf{Q}_{C,i}$ that connect
the magnetic hot spots of the Fermi surface \cite{Comin14}, according
to Fig. \ref{FigScatteringMechanisms}. We note however that small
changes in the positions of the CDW hot spots do not affect our conclusions.
Because $\mathbf{Q}_{C,i}$ and $\mathbf{Q}_{S}$ connect states at
the Fermi level, the CDW and SDW dynamics are dominated by Landau
damping, i.e. $\chi_{\alpha}^{-1}\left(\mathbf{q},\omega\right)=\chi_{\alpha}^{-1}\left(\mathbf{q}\right)-i\omega/\Gamma_{\alpha}$
and $\alpha=C,S$, with $\Gamma_{C/S}\propto v_{F}Q_{C/S}$, where
$v_{F}$ is the Fermi velocity. The anisotropy of the fluctuations
is manifested in their static components, which, according to the
experimental observations, can be modeled as: 
\begin{align}
\chi_{C,i}^{-1}\left(\mathbf{q}+\mathbf{Q}_{C,i}\right) & =\xi_{C}^{-2}\left(1\pm\eta_{C}\right)+q^{2}\label{EqAniSuscept}\\
\chi_{S}^{-1}\left(\mathbf{q}+\mathbf{Q}_{S}\right) & =\xi_{S}^{-2}+\left(1+\eta_{S}\right)q_{x}^{2}+\left(1-\eta_{S}\right)q_{y}^{2}
\end{align}
where the upper (lower) sign in the first equation refers to $i=a$
($i=b$). Hereafter, $\hat{\mathbf{x}}\parallel\hat{\mathbf{a}}$,
$\hat{\mathbf{y}}\parallel\hat{\mathbf{b}}$, and all lengths are
measured in units of the lattice constant. Fig. \ref{FigScatteringMechanisms}
displays the contour plots of the susceptibilities, highlighting their
anisotropic features: while the anisotropy of the CDW fluctuations
is manifested as different correlation lengths \cite{Chubukov14,Norman14,Tsvelik_Chubukov},
$\eta_{C}=\left(\xi_{C,a}^{-2}-\xi_{C,b}^{-2}\right)/2\xi_{C}^{-2}$,
the anisotropy of the SDW fluctuations is manifested only on its form
factor via the dimensionless parameter $\eta_{S}$. When $\left|\eta_{S}\right|>1$,
the SDW develops an incommensurability along either $a$ ($\eta_{S}<0$)
or $b$ ($\eta_{S}>0$). Thus, both $\eta_{S}$ and $\eta_{C}$ are
Ising-nematic order parameters and the anisotropic resistivity obeys,
by symmetry, $\Delta\rho=C_{S}\eta_{S}+C_{C}\eta_{C}$. Because our
main goal is to establish the sign of the pre-factors $C_{S}$ and
$C_{C}$, hereafter we consider the regime $\eta_{S,C}\ll1$.

Because macroscopic samples will be divided in equal-weight domains
of $\eta_{S,C}$ and $-\eta_{S,C}$, one would not expect to observe
anisotropic properties which average over the entire sample, such
as $\Delta\rho$. This issue can be avoided if fields that explicitly
break the tetragonal symmetry and select one domain over the other
are present. In terms of a Ginzburg-Landau functional, they can be
recast in terms of the conjugate fields $h_{C}$ and $h_{S}$: 
\begin{equation}
F[\eta_{\mathrm{S}},\eta_{\mathrm{C}}]=F_{0}[\eta_{\mathrm{S}},\eta_{\mathrm{C}}]-h_{\mathrm{C}}\eta_{\mathrm{C}}-h_{\mathrm{S}}\eta_{\mathrm{S}}\label{EqNematicCoupling}
\end{equation}
where the functional $F_{0}$ depends only on even powers of $\eta_{S,C}^{2}$
and $\eta_{S}\eta_{C}$. In tetragonal cuprates such as HgBa$_{2}$CuO$_{4}$
and Nd$_{2}$CuO$_{4}$ the symmetry-breaking field needs to be externally
applied in the form of uniaxial strain. However, in detwinned YBCO,
the presence of unidirectional CuO chains makes it orthorhombic, with
the $b$ direction parallel to the CuO chains \cite{Atkinson99,Das12}.
Thus, the small orthorhombic distortion acts effectively as an external
field that selects one type of domain \cite{Fernandes14}.

\begin{figure}
\includegraphics[width=0.9\columnwidth]{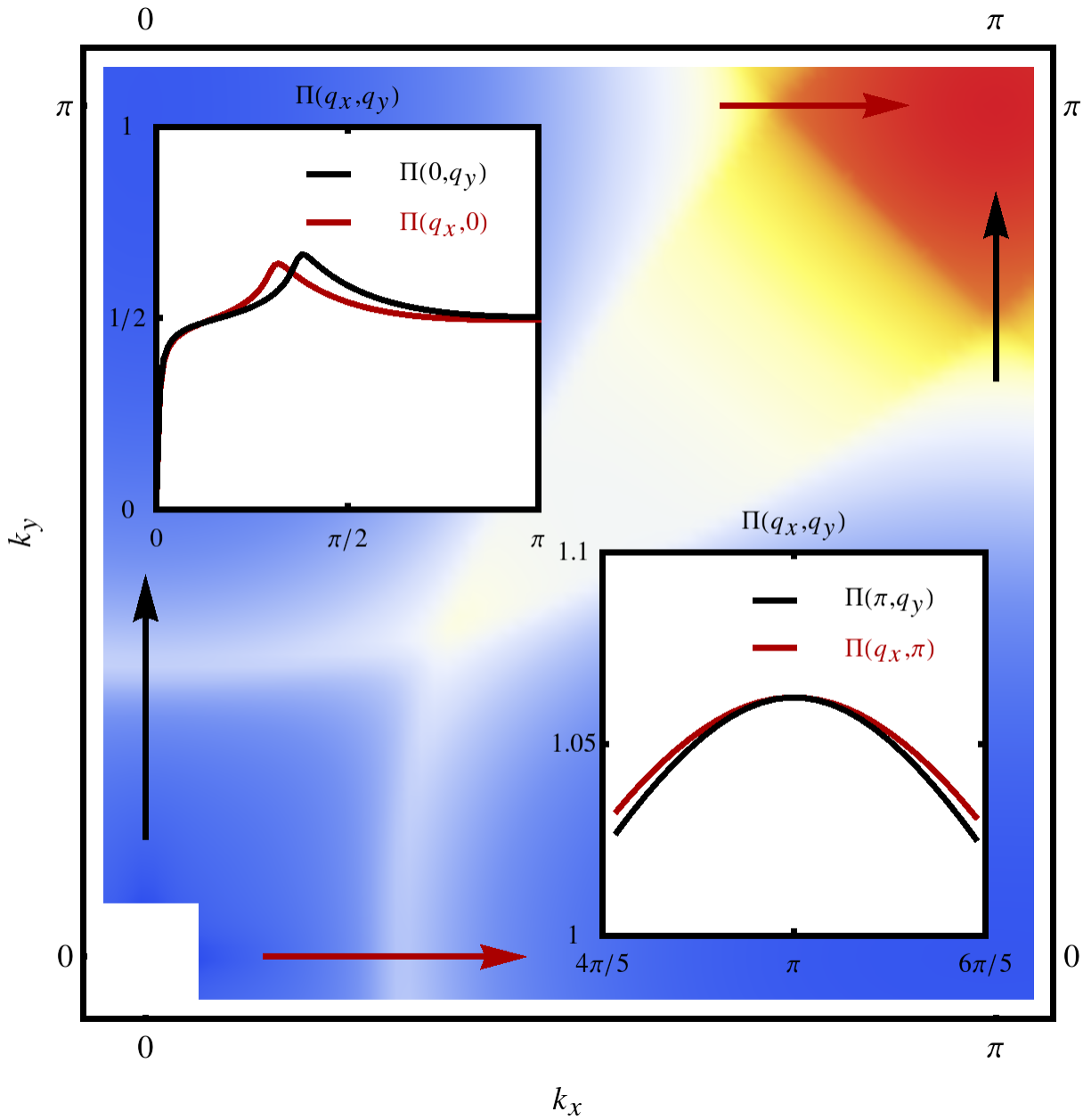} \protect\protect\protect\protect\protect\caption{(color online) Color plot of the polarization bubble $\Pi\left(\mathbf{q}\right)$
across the first Brillouin zone in the presence of a non-zero coupling
between the CuO chain and the CuO$_{2}$ plane. The insets show the
high-symmetry cuts, indicated by the arrows, near the CDW ordering
vectors ($\Pi\left(q_{x},0\right)$ and $\Pi\left(0,q_{y}\right)$),
and near the SDW ordering vector ($\Pi\left(\pi,q_{y}\right)$ and
$\Pi\left(q_{x},\pi\right)$).}

\label{FigDensity} 
\end{figure}

To verify whether this picture correctly captures the signs of $\eta_{S}$
and $\eta_{C}$ observed experimentally in YBCO, namely $\eta_{S}<0$
and $\eta_{C}>0$, we computed the signs of the effective fields $h_{C,S}$
generated by the coupling between the CuO chains and the CuO$_{2}$
planes via evaluation of the non-interacting polarization bubble $\Pi(\mathbf{q},\omega)$
for a tight-binding model containing the chains and the planes \cite{Atkinson99,Das12}
(see supplementary material\footnote{See Supplemental Material [url], which includes Refs.\cite{Das12,Sachdev_LaPlaca,Atkinson99,ziman2001electrons,Rosch99,Fernandes11}}). Because the contribution of the chains
to the susceptibilities (\ref{EqAniSuscept}) is given by $\tilde{\chi}_{\alpha}^{-1}\left(\mathbf{q}\right)-\chi_{\alpha}^{-1}\left(\mathbf{q}\right)=-\Pi\left(\mathbf{q}\right)$,
where $\tilde{\chi}$ is the susceptibility in the presence of the
conjugate fields induced by the chains, it is straightforward to extract
the fields $h_{C,S}$. In Fig. \ref{FigDensity} we plot $\Pi\left(\mathbf{q}\right)$
across the first Brillouin zone, and present in the inset cuts along
the high-symmetry directions $\left(q_{x},0\right)$, $\left(0,q_{y}\right)$,
$\left(\pi+q_{x},\pi\right)$ and $\left(\pi,\pi+q_{y}\right)$.

First, we note that the peaks along the $90^{\circ}$-related cuts
$\Pi\left(q_{x},0\right)$ and $\Pi\left(0,q_{y}\right)$ are different,
with the peak along the $q_{y}$ axis (parallel to $b$) stronger,
which corresponds to a larger correlation length around the $\mathbf{Q}_{C,b}$
ordering vector, $\xi_{C,b}>\xi_{C,a}$. Therefore, the effect of
the chains can be recast in terms of a positive conjugate field $h_{C}>0$
that selects the $\eta_{C}>0$ domain, in agreement with the x-ray
observations in YBCO \cite{Hayden_PRB14,LeTacon14}. Meanwhile, a
cut along the $a$ and $b$ axes centered at the $\mathbf{Q}_{S}=\left(\pi,\pi\right)$
ordering vector gives $\Pi\left(\pi+q_{x},\pi\right)-\Pi\left(\pi,\pi\right)=-\alpha_{x}q_{x}^{2}$
and $\Pi\left(\pi,\pi+q_{y}\right)-\Pi\left(\pi,\pi\right)=-\alpha_{y}q_{y}^{2}$,
with $\alpha_{x}<\alpha_{y}$. Thus, comparison with Eqs. (\ref{EqAniSuscept})
reveals that the chains act as a negative conjugate field $h_{S}<0$,
which selects the $\eta_{S}<0$ domain, as also observed experimentally
in YBCO via neutron scattering \cite{Hinkov08,Keimer_NJP_10}. Note
that, as pointed out in Ref. \cite{Ando02}, even though the chains
contribute to $\Delta\rho$, they cannot alone explain the resistivity
anisotropy behavior, since $\Delta\rho$ has a non-monotonic variation
as doping decreases, whereas the degree of chain order decreases continuously
with decreasing $p$.

Having established the form of the anisotropic SDW and CDW susceptibilities,
we now compute the resistivity anisotropy arising from the scattering
of electrons by these fluctuations. Because we focus on the sign of
$\Delta\rho$ for small $\eta_{C,S}$, it is appropriate to employ
a semi-classical Boltzmann approach \cite{Rosch99,Fernandes11,Rice13},
since the smallness of $\eta_{C,S}$ allows for a perturbative treatment
of the collision kernel, even if the SDW and CDW coupling constants
are not necessarily small. Furthermore, the observations of quantum
oscillations \cite{Taillefer10}, of a $T^{2}$ behavior in the resistivity
\cite{Greven_PNAS}, of the validity of Kohler's rule \cite{Greven_Kohler},
and of a $\omega^{2}$ behavior in the ac conductivity \cite{vanderMarel13}
suggest that quasi-particles are well-defined in the doping range
of interest. We emphasize that our focus is in the underdoped regime
where $\xi_{S,C}$ remains finite, and the system is near a finite-temperature
nematic phase transition. Near a putative nematic quantum critical
point, the quasi-particle concept is compromised, and other approaches
may be more appropriate~\cite{LawlerPRB2006,LawlerPRB2007,NilssonPRB2005}.

Besides the inelastic scattering by CDW and SDW fluctuations, electrons
are also scattered elastically by impurities (see also Refs. \cite{Carlson06,Hirschfeld12}).
Here, we consider the limit where the impurity potential provides
the dominant scattering mechanism, which is always true at low enough
temperatures. Alternatively, similar results can be obtained in the
limit where scattering by isotropic fluctuations is dominant. We avoid
the extremely low-temperature regime, where weak-localization and
Fermi-velocity renormalization effects may be important. In the impurity-dominated
regime \cite{Rosch99,Fernandes11}, the solution of the Boltzmann
equation yields the resistivity anisotropy (see supplementary material):
\begin{equation}
\rho_{a}-\rho_{b}=\rho_{0}\frac{\sum_{\alpha}\left(I_{\mathrm{fluct}}^{\alpha}[h_{x}/\tau]-I_{\mathrm{fluct}}^{\alpha}[h_{y}/\tau]\right)}{I_{\mathrm{imp}}[h/\tau]}\label{EqFunctionalResistiviyAinsotropy}
\end{equation}
with the collision integrals: 
\begin{equation}
I\left[h_{j}\right]=\frac{1}{2\hbar}\int_{\mathbf{p},\mathbf{p}'}\mathcal{K}\left(\mathbf{p},\mathbf{p}'\right)\left(h_{j}(\mathbf{p})-h_{j}(\mathbf{p}')\right)^{2}\label{aux_collision_int}
\end{equation}
and the kernels: 
\begin{eqnarray}
\mathcal{K}_{\mathrm{imp}}\left(\mathbf{p},\mathbf{p}'\right) & = & \frac{g_{0}^{2}}{\beta}\delta(\varepsilon_{\mathbf{p}}-\mu)\delta(\varepsilon_{\mathbf{p}}-\varepsilon_{\mathbf{p}'})\nonumber \\
\mathcal{K}_{\mathrm{fluct}}^{\alpha}\left(\mathbf{p},\mathbf{p}'\right) & = & \frac{g_{\alpha}^{2}}{8}\frac{\sinh[\frac{\beta}{2}(\varepsilon_{\mathbf{p}'}-\varepsilon_{\mathbf{p}})]^{-1}\Im\chi_{\alpha}(\mathbf{p},\mathbf{p}')}{\cosh[\frac{\beta}{2}(\varepsilon_{\mathbf{p}}-\mu)]\cosh[\frac{\beta}{2}(\varepsilon_{\mathbf{p}'}-\mu)]}\label{collision_int}
\end{eqnarray}
Here, $\alpha=C_{a},C_{b},S$ refers to the CDW fluctuations around
the ordering vectors $\mathbf{Q}_{C,a/b}$ and to the SDW fluctuations
around $\mathbf{Q}_{S}$. $h_{j}=\frac{\tau e\beta}{\hbar}\,\frac{\partial\varepsilon_{\mathbf{k}}}{\partial k_{j}}$,
with $i=x,y$, denotes the deviation of the electronic distribution
function $n_{F}$ from the equilibrium Fermi-Dirac distribution $n_{F}^{0}$
in the presence of an electric field $\mathbf{E}$, $n_{F}=n_{F}^{0}-\beta^{-1}\left(\partial_{\epsilon}n_{F}^{0}\right)\mathbf{h}\cdot\mathbf{E}$,
$\tau^{-1}=g_{0}^{2}/(\pi\nu_{F}\hbar)$ is the impurity scattering
rate and $\rho_{0}=\frac{\hbar}{e^{2}}\frac{2\pi}{\hbar\nu_{F}\tau}\frac{1}{\braket{v_{j}^{2}}_{\hat{k}}}$
is the impurity-induced residual resistivity. The electronic dispersion
is denoted by $\varepsilon_{\mathbf{p}}$, the CDW and SDW susceptibilities
$\chi_{\alpha}$ are given by Eq. (\ref{EqAniSuscept}) and $g_{0}$,
$g_{\alpha}$ denote the scattering amplitudes for impurities and
fluctuations, respectively.

\begin{figure}
\includegraphics[width=0.9\columnwidth]{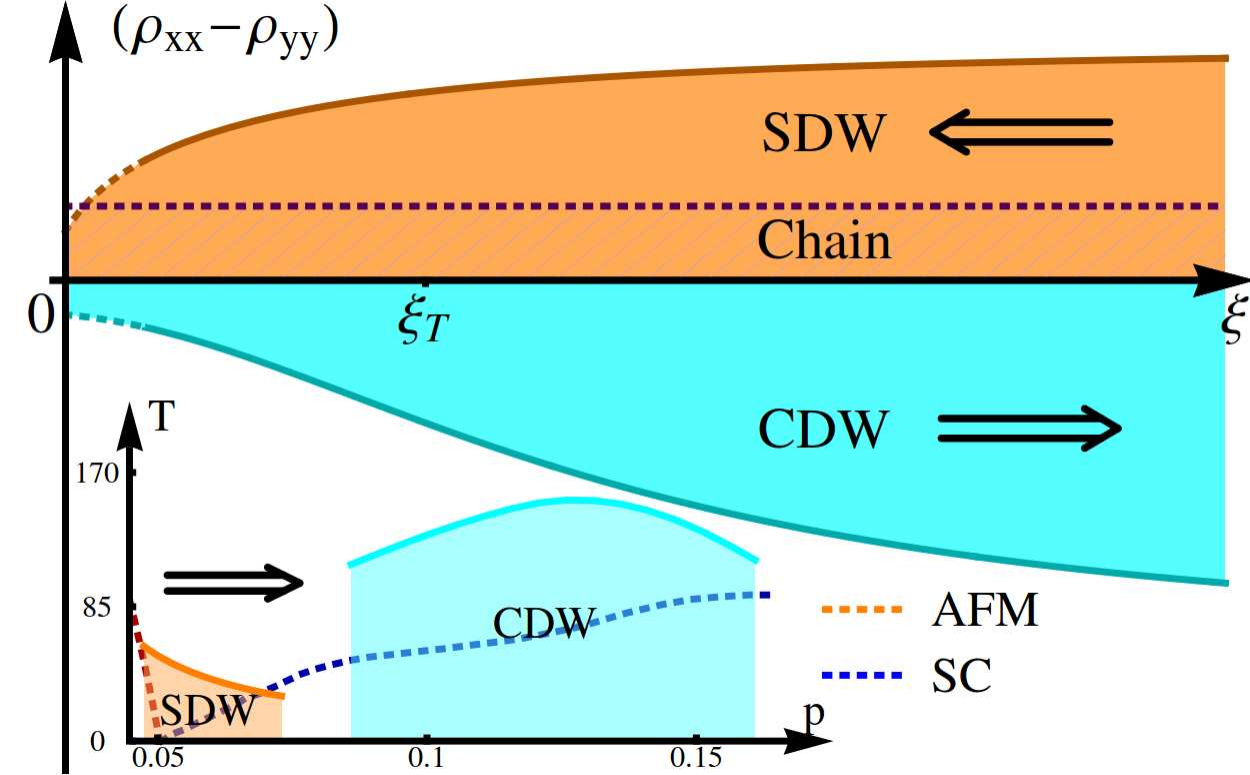}
\protect\protect\protect\protect\protect\caption{(color online) Resistivity anisotropy $\rho_{a}-\rho_{b}$ due to
SDW and CDW fluctuations as function of their correlation lengths
$\xi_{S,C}$. The arrows denote how the correlation lengths change
as doping increases, as shown schematically in the inset. $\xi_{T}\propto\sqrt{\Gamma/T}$
is the length scale associated with the thermal excitations of the
fluctuations. A constant contribution from the CuO chains in YBCO
is indicated as a dashed line.}

\label{FigResistivity} 
\end{figure}

The collision integrals that determine the resistivity anisotropy
(\ref{EqFunctionalResistiviyAinsotropy}) are dominated by their behavior
near the CDW/SDW hot spots, $\varepsilon_{\mathbf{p}+\mathbf{Q}_{\alpha}}=\varepsilon_{\mathbf{p}}=0$,
where the susceptibility $\chi_{\alpha}$ is the largest. For the
CDW fluctuations, Eq. (\ref{EqAniSuscept}), because the anisotropy
is manifested in the correlation length we find that the anisotropy
depends only on the Fermi velocity at the hot spots. Introducing the
average distance between thermally induced fluctuations $\xi_{T}=\sqrt{\frac{3\Gamma_{C}\beta}{2\pi}}$,
we obtain in the low-temperature limit $\xi_{T}\gg\xi_{C}\gg1$ the
leading-order expression: 
\begin{equation}
\left(\frac{\rho_{a}-\rho_{b}}{\rho_{0}}\right)_{\mathrm{C}}\approx\left(\frac{g_{C}^{2}\xi_{C}^{2}}{g_{0}^{2}\beta\chi_{0,C}^{-1}\xi_{T}^{2}}\right)C_{C}\,\eta_{C}\label{EqCDWResistivityResult}
\end{equation}
where $\chi_{0,C}^{-1}$ is the CDW energy scale and $C_{C}<0$ is
a dimensionless positive constant that depends only on the Fermi velocity
at the CDW hot spots. Therefore, in YBCO, since $\eta_{C}>0$, scattering
by charge fluctuations favor $\rho_{a}<\rho_{b}$. This can be understood
in the following way: since $\eta_{C}>0$, fluctuations are stronger
around the $\mathbf{Q}_{C,b}$ CDW ordering vector, i.e. $\xi_{C,b}>\xi_{C,a}$.
As shown in Fig. \ref{FigScatteringMechanisms}, at the hot spots
connected by $\mathbf{Q}_{C,b}$, the Fermi velocity is almost parallel
to the $b$ axis. Thus, electrons moving along the $b$ direction
experience enhanced scattering compared to the electrons moving along
$a$, causing $\rho_{a}<\rho_{b}$. This argument makes it clear that
small deviations in the value of $Q_{C}$ do not change the result.

As for the SDW fluctuations, the anisotropy does not arise from the
ordering vector $\mathbf{Q}_{S}=(\pi,\pi)$, which is isotropic, but
from the form factor. As a result, defining again $\xi_{T}=\sqrt{\frac{3\Gamma_{S}\beta}{2\pi}}$
and focusing in the regime $\xi_{T}\gg\xi_{S}\gg1$, we obtain: 
\begin{equation}
\left(\frac{\rho_{a}-\rho_{b}}{\rho_{0}}\right)_{\mathrm{S}}\approx\left(\frac{g_{S}^{2}\ln\xi_{S}}{g_{0}^{2}\beta\chi_{0,S}^{-1}\xi_{T}^{2}}\right)C_{S}\,\eta_{S}\label{EqSDWResistivityResult}
\end{equation}
In contrast to the CDW case, the dimensionless pre-factor $C_{\mathrm{S}}$
depends on the curvature of the Fermi surface and on the derivatives
of the Fermi velocity near the hot spots. As a result, $C_{S}$ may
depend on additional details of the Fermi surface, as compared to
$C_{C}$. We computed it using two different sets of tight-binding
parameters \cite{Das12,Sachdev_LaPlaca} and different values of the
chemical potential, finding that in general $C_{S}<0$. Consequently,
since $\eta_{S}<0$ in YBCO, scattering by SDW fluctuations yields
$\rho_{a}>\rho_{b}$. This can be understood as a consequence of the
fact that the SDW fluctuations stiffness is smaller along the $a$
axis, since $\eta_{S}<0$ in Eq. (\ref{EqAniSuscept}), which enhances
the scattering along this direction. Note that, because long-range
SDW order is present while long-range CDW order is absent in the underdoped
phase diagram, $\xi_{S}$ can become very large whereas $\xi_{C}$
remains bounded.

We now contrast our results to the experimental measurements of $\Delta\rho\equiv\left(\rho_{a}-\rho_{b}\right)/\rho_{b}$
\cite{Ando02}. In YBCO, the CuO chains, parallel to the $b$ axis,
give an intrinsic contribution to the resistivity anisotropy, $\Delta\rho_{\mathrm{chain}}>0$
(see dashed line in Fig. \ref{FigResistivity}). Thus, the contribution
from the CDW/SDW fluctuations add to or subtract from this intrinsic
background. As shown in the inset of Fig. \ref{FigResistivity}, anisotropic
SDW and CDW fluctuations compete and dominate different regions of
the underdoped phase diagram. Starting at $p\approx0.05$ and increasing
$p$, the anisotropic SDW fluctuations with $\eta_{S}<0$ are suppressed
as the corresponding transition line disappears near $p\approx0.08$
\cite{LeTacon14,Hayden_PRB14}. According to our results, $\Delta\rho$
should be positive and should decrease as $p$ increases and $\xi_{S}$
is suppressed, as shown by the arrow in Fig. \ref{FigResistivity}.
This behavior is indeed observed experimentally \cite{Ando02}. CDW
fluctuations emerge at $p\approx0.09$ -- initially they are anisotropic,
with $\eta_{C}>0$, but as $p\approx0.13$ is approached they become
isotropic \cite{LeTacon14}, with $\eta_{C}\rightarrow0$. In this
regime, we find that the anisotropic CDW fluctuations give $\Delta\rho<0$.
Experimentally, the measured $\Delta\rho$ remains positive in this
region, but is the smallest in the phase diagram \cite{Ando02}, which
could be understood as a consequence of $\Delta\rho<0$ appearing
on the intrisinc $\Delta\rho_{\mathrm{chain}}>0$ background. To shed
light on this issue and disentangle the chains contribution, it would
be desirable to perform transport measurements in tetragonal compounds
such as HgBa$_{2}$CuO$_{4}$ and Nd$_{2}$CuO$_{4}$, where CDW fluctuations
have also been reported \cite{Damascelli14,Greven_CDW}. In this case,
application of uniaxial strain \cite{Fisher10,Tanatar10} would be
necessary to select a single nematic domain. Note that for very underdoped
YBCO samples, long-range SDW order sets in at very low temperatures
\cite{Keimer_NJP_10}, giving rise to an anisotropic reconstructed
Fermi surface, which promote a non-zero $\Delta\rho$ even in the
absence of inelastic scattering at $T=0$.

In summary, we have shown that the anisotropic charge and spin fluctuations
present in YBCO give antagonistic contributions to the resistivity
anisotropy in underdoped cuprates. While the SDW fluctuations provide
a plausible explanation for the resistivity anisotropy observed experimentally,
the contribution of CDW fluctuations seems to be nearly cancelled
by the contribution coming from the CuO chains. An open issue is how
these anisotropic fluctuations affect other anisotropic transport
quantities, such as the thermopower and the Nernst anisotropy \cite{Taillefer10}.
Although a non-zero $\Delta\rho$ is not surprising, since these fluctuations
are $C_{2}$ symmetric, the fact that the competing fluctuating channels
promote different signs for $\Delta\rho$ is unanticipated, opening
a promising route to disentangle the contributions from spin and charge
degrees of freedom to the formation of the nematic state observed
in underdoped cuprates. 

We thank M. Chan, A. Chubukov, M. Greven, M. Le Tacon, and J. Schmalian
for fruitful discussions. MS acknowledges the support from the Humboldt
Foundation. RMF is supported by the U.S. Department of Energy under
Award Number $\text{DE-SC0012336}$.

\bibliographystyle{apsrev4-1}
\bibliography{bibliography}

\begin{widetext}
\newpage{}

\section{supplementary material}

\subsection{Anisotropic fluctuations and the coupling to the chains in YBCO}

As explained in the main text, the changes in the susceptibility caused
by the coupling to the CuO chains present in the YBCO compounds can
be evaluated via the polarization operator: 
\begin{equation}
\tilde{\chi}_{\alpha}^{-1}\left(\mathbf{q}\right)-\left[\chi_{\alpha}^{-1}\left(\mathbf{q}\right)\right]_{\eta_{C}=\eta_{S}=0}=-\delta\Pi\left(\mathbf{q}\right).
\end{equation}
with $\alpha=C,S$ and $\chi_{\alpha}^{-1}$ given by Eq. (1) of the
main text. Here we are interested only in the anisotropic properties:
$\delta\Pi\left(\mathbf{q}\right)=\Pi\left(\mathbf{q}\right)-\Pi_{\mathrm{no-chain}}\left(\mathbf{q}\right)$,
which by definition must arise from the coupling to the CuO chains,
since the electronic dispersion due to the CuO$_{2}$ planes is tetragonally
symmetric. We emphasize that, in our approach, the role played by
the chains is to simply induce a conjugate field that selects a particular
nematic domain, and not to cause the nematic instability in the first
place. The polarization operator is given by the standard expression:

\begin{equation}
\Pi(\mathbf{q})=\beta^{-1}\sum_{\omega_{n}}\int\frac{\mathrm{d}^{2}p}{(2\pi)^{2}}\mathrm{Tr}\left[\mathcal{G}(i\omega_{n},\mathbf{p})\mathcal{G}(i\omega_{n},\mathbf{p}-\mathbf{q})\right],
\end{equation}
where $\mathcal{G}$ denotes the non-interacting matrix Green's function
of the multi-band system consisting of the CuO$_{2}$ plane and the
CuO chain. Even though YBCO has two CuO$_{2}$ planes per unit cell,
the main results are captured by a simpler two-band model consisting
of a single plane and a single chain, defined via the spinor $\Psi_{\mathbf{k}\sigma}^{\dagger}=\left(\begin{array}{cc}
d_{p,\mathbf{k}\sigma}^{\dagger} & d_{c,\mathbf{k}\sigma}^{\dagger}\end{array}\right)$ where $p,c$ denote plane or chain operators, respectively. The corresponding
non-interacting Hamiltonian is therefore given by $H^{(2)}=\sum_{\mathbf{k}\sigma}\Psi_{\mathbf{k}\sigma}^{\dagger}\mathcal{H}_{\mathbf{k}}^{(2)}\Psi_{\mathbf{k}\sigma}$
with the matrix:

\begin{equation}
\mathcal{H}_{\mathbf{k}}^{(2)}=\begin{pmatrix}\epsilon_{p}(\vec{k}) & t_{cp}\\
t_{cp} & \epsilon_{c}(\vec{k})
\end{pmatrix}
\end{equation}

Here we defined the tight-binding dispersions of the plane and of
the chain \cite{Das12,Sachdev_LaPlaca}:

\begin{align}
\epsilon_{p}(\vec{k}) & =-2t(\cos k_{x}+\cos k_{y})-4t'\cos k_{x}\cos k_{y}-2t''(\cos2k_{x}+\cos2k_{y})-\mu_{p},\nonumber \\
\epsilon_{c}(\vec{k}) & =-2t_{c}\cos k_{y}-\mu_{c}.
\end{align}
and the plane-chain hopping parameter $t_{cp}$. The main effects
of the coupling to the chain can be understood analytically by considering
the limit $t_{cp}\ll\left|\mu_{c}\right|$. In this case, the effective
plane dispersion becomes:

\begin{equation}
\tilde{\epsilon}_{p}(\vec{k})=-2\tilde{t}(\cos k_{x}+\cos k_{y})+2\delta(\cos k_{x}-\cos k_{y})-4t'\cos k_{x}\cos k_{y}-2t''(\cos2k_{x}+\cos2k_{y})-\tilde{\mu}_{p}\label{renormalized_band}
\end{equation}
with the modified tight-binding parameters: $\tilde{t}=t-\left(\frac{t_{cp}}{\mu_{c}}\right)^{2}(t-t_{c})$,
$\delta=\left(\frac{t_{cp}}{\mu_{c}}\right)^{2}t_{c}$ and $\tilde{\mu}_{p}=\mu_{p}+\frac{2t_{cp}^{2}}{|\mu_{c}|}$.
The main change in the dispersion is the appearance of the anisotropic
term with coefficient $\delta>0$. Since $\tilde{t}>0$, this term
effectively reduces the Fermi-momentum along the $k_{y}$ direction.
As a result, the Fermi surface is squeezed (relative to the $\left(\pi,\pi\right)$
point), becoming more elongated along the $k_{x}$ axis than along
the $k_{y}$ axis, as shown in Figure \ref{fig_FS}.

\begin{figure}
\begin{centering}
\includegraphics[width=0.35\columnwidth]{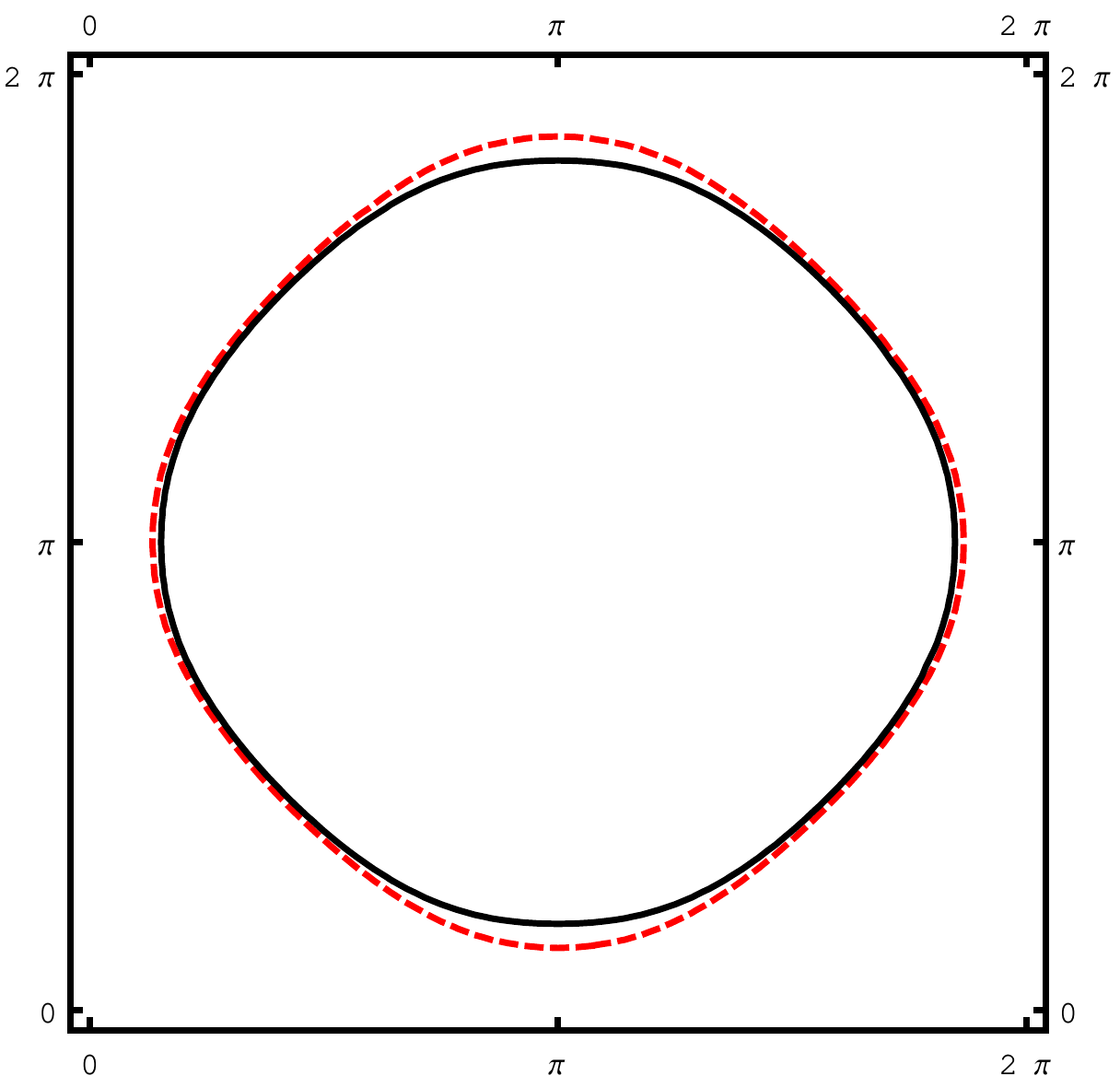} 
\par\end{centering}

\protect\protect\protect\caption{Distortion of the Fermi surface due to the coupling between the plane
and the CuO chain. For clarity, we increase the value of $t_{cp}$
to $0.1$, we shift the origin of the Brillouin zone to $\left(\pi,\pi\right)$
and display both the distorted (solid line) and undistorted (dashed
line) Fermi surfaces. \label{fig_FS}}
\end{figure}

The impact of these changes on the polarization operator can be understood
in a straightforward way. The squeezing of the Fermi surface promotes
extended and relatively flat segments displaced along the $k_{y}$
direction. Because these flat segments provide an enhanced contribution
to the density response, the CDW fluctuations near the ordering vector
$\mathbf{Q}_{C,b}$ are favored compared to the fluctuations centered
at $\mathbf{Q}_{C,a}$. Similarly, the SDW fluctuations, centered
at $\mathbf{Q}_{S}=\left(\pi,\pi\right)$, become stiffer along the
$k_{y}$ direction, as compared to the $k_{x}$ direction.

For the numerical evaluation presented in Figure 2 of the main text,
we used the band structure parameters of Ref. \cite{Das12}. In particular,
to make the chain effect more visible, we considered a larger value
of the chain-plane coupling $t_{cp}$ than the one in Ref. \cite{Das12}.
The parameters used were, in eV, $(t,t_{1},t_{2},\mu_{p},t_{c},\mu_{c},t_{cp})==(0.38,0.0684,0.095,0,0.25,-0.87,-0.075)$.
As mentioned above, in YBCO two planes are actually coupled to the
same CuO chain in each unit cell. A more precise model therefore starts
with the three-component spinor $\Psi_{\mathbf{k}\sigma}^{\dagger}=\left(\begin{array}{ccc}
d_{p_{1},\mathbf{k}\sigma}^{\dagger} & d_{p_{2},\mathbf{k}\sigma}^{\dagger} & d_{c,\mathbf{k}\sigma}^{\dagger}\end{array}\right)$ and the Hamiltonian $H^{(3)}=\sum_{\mathbf{k}\sigma}\Psi_{\mathbf{k}\sigma}^{\dagger}\mathcal{H}_{\mathbf{k}}^{(3)}\Psi_{\mathbf{k}\sigma}$
with \cite{Atkinson99}: 
\begin{equation}
\mathcal{H}_{\mathbf{k}}^{(3)}=\begin{pmatrix}\epsilon_{p}(\vec{k}) & t_{pp} & t_{cp}\\
\epsilon_{pp} & \epsilon_{p}(\vec{k}) & t_{cp}\\
t_{cp} & t_{cp} & \epsilon_{c}(\vec{k})
\end{pmatrix}
\end{equation}

The non-zero inter-plane hopping $t_{pp}>0$ gives rise to bonding
and anti-bonding bands, i.e. $\epsilon_{p}^{(\pm)}(\vec{k})=\epsilon_{p}(\vec{k})\pm t_{pp}$.
Once the coupling to the chains is included, only the anti-bonding
band $\epsilon_{p}^{(+)}(\vec{k})$ is in fact affected by $t_{cp}$
and becomes anisotropic, similarly to Eq. (\ref{renormalized_band}).
Therefore, as long as the plane-chain coupling is not too large compared
to the inter-plane hopping, $\left|t_{cp}\right|\ll t_{pp}$, we find
that the anisotropy in the polarization operator is the same as in
the case of a single plane coupled to the chains, since the anisotropy
of the Fermi surface is the same as in Eq. (\ref{renormalized_band}).

\subsection{Anisotropic Transport: Boltzmann equation formalism}

\subsubsection{Fermi surface parametrization}

To solve the Boltzmann equation, it is convenient to have a suitable
parametrization of the Fermi surface of the usual cuprate tight-binding
models \cite{Das12,Sachdev_LaPlaca}. Since we are interested in the
hole-underdoped regime, it is convenient to shift the center of the
Brillouin zone to $(\pi,\pi)$, around which the Fermi surface is
closed. In particular, it can be parametrized by: 
\begin{equation}
\vec{p}_{F}(\phi)=\bar{p}_{F}f(\phi)\begin{pmatrix}\cos\phi\\
\sin\phi
\end{pmatrix}\quad\text{while }\epsilon_{p}(\vec{p}_{F}(\phi))-\mu=0,
\end{equation}
where $f$ is a dimensionless function encoding the form of the Fermi
surface and $\bar{p}_{F}$ is a radial coordinate proportional to
the Fermi momentum scale. Here, $\phi$ is the angle measured relative
to the $k_{x}$ axis. Away from the Fermi level, the momentum is parametrized
by $\vec{p}(\bar{p},\phi)=\vec{p}_{F}(\phi)\bar{p}/\bar{p}_{F}$.
Thus, near the Fermi level, we can expand the dispersion as: 
\begin{equation}
\epsilon_{p}(\vec{p}(\bar{p},\phi))-\mu\approx
\vec{v}_{F}(\phi)\cdot\vec{p}_{F}(\phi)(\frac{\bar{p}}{\bar{p}_{F}}-1)
\end{equation}
allowing us to express the radial component in terms of an energy
variable $\epsilon$, $\bar{p}/\bar{p}_{F}=1+\epsilon/(\vec{v}_{F}(\phi)\cdot\vec{p}_{F}(\phi))$.
For convenience, we identify the angle-dependent energy scale $\vec{v}_{F}(\phi)\cdot\vec{p}_{F}(\phi)=\epsilon_{F}(\phi)$.
To evaluate the sums over momentum that appear in the Boltzmann equation
solution, we define the angle-dependent density of states $N_{\phi}=\vec{p}_{F}^{2}(\phi)/\left|\epsilon_{F}(\phi)\right|$,
such that the total density of states $\nu_{F}$ is given by $\nu_{F}=\int\frac{\mathrm{d}\phi}{2\pi}N_{\phi}$.
Then, for an arbitrary function $K$ strongly peaked at the Fermi
surface we have: 
\begin{equation}
\int\frac{\mathrm{d}^{2}p}{(2\pi\hbar)^{2}}K(\epsilon_{p}(\vec{p})-\mu,\vec{p})\approx\nu_{F}\braket{\int\frac{\mathrm{d}\epsilon}{2\pi}K(\epsilon,\vec{p}_{F}(\phi))}_{\phi}.
\end{equation}
where we introduced the notation:

\begin{equation}
\nu_{F}\braket{F(\phi)}_{\phi}=\int\frac{\mathrm{d}\phi}{2\pi}N_{\phi}F(\phi).
\end{equation}

\subsubsection{Functional Approach to the Boltzmann equation}

The Boltzmann equation for scattering by impurities and fluctuations
is given by:

\begin{equation}
\left(\pd{}{t}+\vec{v}\cdot\nabla_{\vec{r}}+e\vec{E}\cdot\vec{v}\pd{}{\epsilon}\right)n_{F}=-\mathrm{\mathcal{I}}_{\mathrm{imp}}[n_{F}]-\sum_{\substack{\alpha}
}\mathrm{\mathcal{I}}_{\mathrm{fluct}}^{\alpha}[n_{F}],
\end{equation}
with $\alpha=C_{a},C_{b},S$. In linear response, we consider weak
perturbations around equilibrium: $n_{F}=n_{F}^{0}-\pd{n_{F}^{0}}{\epsilon}\beta^{-1}h$.
Keeping only the driving field, the linearized Boltzmann equation
becomes: 
\begin{equation}
e\vec{E}\cdot\vec{v}\pd{n_{F}^{0}}{\epsilon}=-\delta\mathcal{I}_{\mathrm{imp}}[h]-\sum_{\alpha}\delta\mathrm{\mathcal{I}}_{\mathrm{fluct}}^{\alpha}[h].\label{EqLinearizedBoltzmanFormal}
\end{equation}

Instead of solving the integral equation above, the solution of the
Boltzmann equation can be obtained by minimization of the functional~\cite{ziman2001electrons,Rosch99,Fernandes11}:
\begin{equation}
F[h]=-D[h]+I_{\mathrm{imp}}[h]+\sum_{\alpha}I_{\mathrm{fluct}}^{\alpha}[h]
\end{equation}

Because $\epsilon_{F}(\phi)\gg T$, we have $\pd{n_{F}^{0}}{\epsilon}\approx-\delta(\epsilon_{F}-\mu)$,
implying that the deviations from equilibrium depend only on the angle
parametrizing the Fermi-surface, $h(\phi)$. Defining the unit vector
$\vec{E}_{i}=E\vec{e}_{i}$, the functionals can be expressed as:

\begin{align}
D_{i}[h_{j}] & =\frac{(eE)\nu_{F}}{2\pi}\braket{\vec{e}_{i}\cdot\vec{v}_{F}(\phi)h_{j}(\phi)}_{\phi},\nonumber \\
I_{\mathrm{imp}}[h_{j}] & =\frac{g_{0}^{2}}{2(2\pi)^{2}\hbar\beta}\braket{\left(h_{j}(\phi)-h_{j}(\phi')\right)^{2}}_{\phi,\phi'},\nonumber \\
I_{\mathrm{fluct}}^{\alpha}[h_{j}] & =\frac{g_{\alpha}^{2}}{16\hbar}\braket{\int\frac{\mathrm{d}\epsilon}{2\pi}\int\frac{\mathrm{d}\epsilon'}{2\pi}\frac{\Im\chi_{\alpha}(\epsilon'-\epsilon;\phi,\phi')\left(h_{j}(\phi)-h_{j}(\phi')\right)^{2}}{\cosh[\frac{\beta}{2}\epsilon]\cosh[\frac{\beta}{2}\epsilon']\sinh[\frac{\beta}{2}(\epsilon'-\epsilon)]}}_{\phi,\phi'}
\end{align}

We are interested in the regime where the elastic impurity scattering
is dominant. In this regime, the deviation from equilibrium is given
by: $h_{i}=\tau\beta eE\vec{e}_{i}\cdot\vec{v}$, where $\tau^{-1}$
is the impurity scattering rate, yielding $D_{i}[\beta e\vec{E}_{j}\cdot\vec{v}]=(eE)^{2}\bar{D}_{ij}$,
$I_{\mathrm{imp}}[\beta e\vec{E}_{j}\cdot\vec{v}]=(eE)^{2}\bar{I}_{\mathrm{eff},j}$
and $I_{\mathrm{fluct}}^{i}[\beta e\vec{E}_{j}\cdot\vec{v}]=(eE)^{2}\bar{I}_{\mathrm{fluct}}^{\alpha}$.
The residual resistivity is therefore given by:

\begin{equation}
\rho_{0}^{-1}=\frac{1}{2}\left(\frac{D_{x}[h_{x}]}{\beta E^{2}}+\frac{D_{y}[h_{y}]}{\beta E^{2}}\right)=\frac{e^{2}}{\hbar}\frac{\hbar^{2}\nu_{F}^{2}}{2g_{0}^{2}}\braket{v_{j}^{2}}_{\phi}=\frac{e^{2}}{\hbar}\frac{\hbar\nu_{F}\tau}{2\pi}\braket{v_{j}^{2}}_{\phi}
\end{equation}

Note that, in principle, since we are interested in the resistivity
anisotropy, we could also include the isotropic contribution from
the fluctuations to the definition of an effective isotropic scattering
rate $\tau^{-1}$. The anisotropic resistivity becomes: 
\begin{equation}
\rho_{a}-\rho_{b}=\rho_{0}\frac{\sum_{\substack{\alpha}
}\delta\bar{I}_{\mathrm{fluct}}^{\alpha}}{\bar{I}_{\mathrm{imp}}}\label{EqFunctionalResistiviyAinsotropySP}
\end{equation}
with: 
\begin{align}
\bar{I}_{\mathrm{imp}} & =\frac{g_{0}^{2}\beta}{(2\pi)^{2}\hbar}\braket{\vec{v}_{F}^{2}(\phi)}_{\phi}\\
\delta\bar{I}_{\mathrm{fluct}}^{\alpha} & =\frac{g_{\alpha}^{2}\chi_{0,\alpha}}{8\pi\hbar}J_{\mathrm{fluct}}^{\alpha}
\end{align}
with $J_{\mathrm{fluct}}^{\alpha}$ given by:

\begin{equation}
J_{\mathrm{fluct}}^{\alpha}=\frac{\pi\beta^{2}}{2\nu_{F}^{2}\chi_{0,\alpha}}\int\frac{\mathrm{d}^{2}p'}{(2\pi\hbar)^{2}}\int\frac{\mathrm{d}^{2}p}{(2\pi\hbar)^{2}}\frac{\left(\vec{e}_{x}\cdot\vec{v}-\vec{e}_{x}\cdot\vec{v}'\right)^{2}-\left(\vec{e}_{y}\cdot\vec{v}-\vec{e}_{y}\cdot\vec{v}'\right)^{2}}{\cosh[\frac{\beta}{2}(\epsilon_{p}-\mu)]\cosh[\frac{\beta}{2}(\epsilon_{p'}-\mu)]\sinh[\frac{\beta}{2}(\epsilon_{p'}-\epsilon_{p})]}\Im\chi_{\alpha}(\epsilon'-\epsilon;\phi,\phi')\label{eq_J}
\end{equation}

Here, we defined the energy scale $\chi_{0,\alpha}^{-1}$ that characterizes
the fluctuation spectrum. The resistivity anisotropy is then given
by: 
\begin{equation}
\rho_{a}-\rho_{b}=\rho_{0}\sum_{\substack{\alpha}
}\frac{\pi}{2}\frac{g_{\alpha}^{2}}{g_{0}^{2}}\frac{\chi_{0,\alpha}}{\beta}\frac{J_{\mathrm{fluct}}^{\alpha}}{\braket{\vec{v}_{F}^{2}(\phi)}_{\phi}}\label{EqResistivityFromJ}
\end{equation}

Note that, in this Boltzmann equation approach, we neglect the renormalization
of the Fermi velocity by the fluctuations as well as weak-localization
corrections. These contributions are only important at very low temperatures,
in the regime $T\ll\tau^{-1}$ , which is not relevant for our analysis.

\subsection{evaluation of the resistivity anisotropy}

\subsubsection{Hot spots contribution}

\label{SubSecCollK} In this section we present the numerical and
analytical evaluation of $J_{\mathrm{fluct}}^{\alpha}$ defined in
Eq. (\ref{eq_J}), obtaining consequently the resistivity anisotropy
(\ref{EqResistivityFromJ}). First, we define the general form of
the susceptibility that can describe either of the CDW fluctuations
(around $\mathbf{Q}_{C,a}$ and $\mathbf{Q}_{C,b}$) or the SDW fluctuations
(around $\mathbf{Q}_{S}$):

\begin{equation}
\frac{\chi_{\alpha}^{-1}(\vec{q},\omega)}{\chi_{0}^{-1}}=\left[\xi_{\alpha}^{-2}+(1+\eta_{\alpha})\left(q_{x}-Q_{\alpha,x}\right)^{2}+(1-\eta_{\alpha})\left(q_{y}-Q_{\alpha,y}\right)^{2}\right]-i\frac{\omega}{\Gamma_{\alpha}}\equiv\tilde{\omega}_{\alpha,q}-i\frac{\omega}{\Gamma_{\alpha}}.\label{EqGeneralSusceptibilityForTransport-1}
\end{equation}
where all lengths are measured relative to the lattice parameter $a$.
Hereafter, for simplicity of notation, we drop the subscript $\alpha$.
Defining the length scale $\xi_{T}^{2}=3\Gamma\beta/(2\pi)$ measuring
the mean distance between thermally excited fluctuations, and using
the following integral evaluation/approximation:

\begin{align}
\int\mathrm{d}\epsilon\frac{1}{\cosh[\frac{\beta}{2}(\epsilon_{p}-\mu)]\cosh[\frac{\beta}{2}(\epsilon_{p}+\omega-\mu)]} & =\frac{2\omega}{\sinh[\omega\beta/2]}\\
\int\mathrm{d}\omega\frac{\omega}{\sinh[\omega\beta/2]^{2}}\frac{\omega\,\chi_{0}\Gamma}{\Gamma^{2}\tilde{\omega}_{q}^{2}+\omega^{2}} & \approx\frac{4\chi_{0}}{3\Gamma\beta^{3}}\frac{2\pi^{2}}{\tilde{\omega}_{q}(\tilde{\omega}_{q}+\frac{2\pi}{3\Gamma\beta})}.
\end{align}
we obtain:

\begin{equation}
J_{\mathrm{fluct}}^{\alpha}=\xi_{T}^{-2}\braket{\smash{\frac{\left(\vec{e}_{x}\cdot\vec{v}-\vec{e}_{x}\cdot\vec{v}'\right)^{2}-\left(\vec{e}_{y}\cdot\vec{v}-\vec{e}_{y}\cdot\vec{v}'\right)^{2}}{\tilde{\omega}_{q}(\phi,\phi')\left[\tilde{\omega}_{q}(\phi,\phi')+\xi_{T}^{-2}\right]}}}_{\phi,\phi'}\label{J_evaluated}
\end{equation}

Clearly, the main contribution to the integral above comes from the
hot spots, which can be parametrized by two angles $\phi_{1}$ and
$\phi_{2}$ defined via $\vec{Q}=\vec{q}=\vec{p}_{F}(\phi_{1})-\vec{p}_{F}(\phi_{2})$.
In the following, we compute Eq. (\ref{J_evaluated}) both numerically,
using the tight-binding dispersion of Ref. \cite{Das12}, and analytically
via an expansion near the hot spots. We consider the CDW and SDW cases
separately, for convenience.

\subsubsection{CDW fluctuations}

The CDW hot spots are connected by the ordering vectors $\mathbf{Q}_{C,a}=Q_{c}\hat{\mathbf{x}}$
and $Q_{C,b}=Q_{c}\hat{\mathbf{y}}$. In the coordinate system centered
at the $\left(\pi,\pi\right)$ point of the Brillouin zone, pairs
of hot spots connected by $\mathbf{Q}_{C,a}$ correspond to $\phi_{1}$,
$\phi_{2}=\pi-\phi_{1}$, whereas the pairs of hot spots connected
by $\mathbf{Q}_{c,b}$ correspond to $\phi_{1}$, $\phi_{2}=-\phi_{1}$.
Expansion around these angles, for the hot spots connected by $\mathbf{Q}_{C,a}$,
gives: 
\begin{align}
q_{y}-Q_{a,y} & \approx\left(\delta\phi_{1}+\delta\phi_{2}\right)\left[\pd{p_{y}}{\phi_{1}}+\frac{1}{2}\pdn{p_{y}}{\phi_{1}}{2}\left(\delta\phi_{1}-\delta\phi_{2}\right)\right]\\
q_{x}-Q_{a,x,} & \approx\left(\delta\phi_{1}-\delta\phi_{2}\right)\left[\pd{p_{x}}{\phi_{1}}+\frac{1}{2}\pdn{p_{x}}{\phi_{1}}{2}\left(\delta\phi_{1}+\delta\phi_{2}\right)\right]
\end{align}

For the hot spots connected by $\mathbf{Q}_{C,b}$, the two functional
forms of the right-hand sides are exchanged. As a result, we obtain:
\begin{equation}
\tilde{\omega}_{q}=\xi_{C,j}^{2}+\left(\pd{p_{x}}{\phi_{1}}\right)^{2}r^{2}+\left(\pd{p_{y}}{\phi_{1}}\right)^{2}s^{2}
\end{equation}
where $r$ and $s$ are defined as $r=\delta\phi_{1}-\delta\phi_{2}$
and $s=\delta\phi_{1}+\delta\phi_{2}$ for $\mathbf{Q}_{C,a}$ (both
are exchanged for $\mathbf{Q}_{C,b}$). Evaluation of the integral
in Eq. (\ref{J_evaluated}), with $J_{\mathrm{fluct}}^{\alpha}=J_{\mathrm{fluct}}^{\alpha,x}+J_{\mathrm{fluct}}^{\alpha,y}$,
gives:

\[
J_{\mathrm{fluct}}^{C,j}=\frac{1}{4\pi}\left(\frac{N_{\phi_{1}}}{\nu_{F}}\right)^{2}\frac{\log\left[1+\left(\frac{\xi_{C,j}}{\xi_{T}}\right)^{2}\right]}{\left|\pd{p_{x}}{\phi_{1}}\pd{p_{y}}{\phi_{1}}\right|}\left\{ \begin{array}{cc}
4v_{x}^{2}(\phi_{1}^{x}) & \mathrm{for}\;\mathbf{Q}_{C,a}\\
-4v_{y}^{2}(\phi_{1}^{y}) & \mathrm{for}\;\mathbf{Q}_{C,b}
\end{array}\right.
\]
where we used $N_{\phi_{1}}=N_{\phi_{2}}$. Because $v_{y}^{2}(\phi_{1}^{y}=\phi_{1}^{x}+\frac{\pi}{2})=v_{x}^{2}(\phi_{1}^{x})$,
the only term that gives rise to an anisotropic resistivity is $\xi_{\mathrm{C},a}\neq\xi_{\mathrm{C},b}$.
Expanding to leading order in the nematic order parameter $\eta_{C}=\left(\xi_{C,a}^{-2}-\xi_{C,b}^{-2}\right)/2\xi_{C}^{-2}$
yields:

\begin{figure}
\includegraphics[width=0.35\columnwidth]{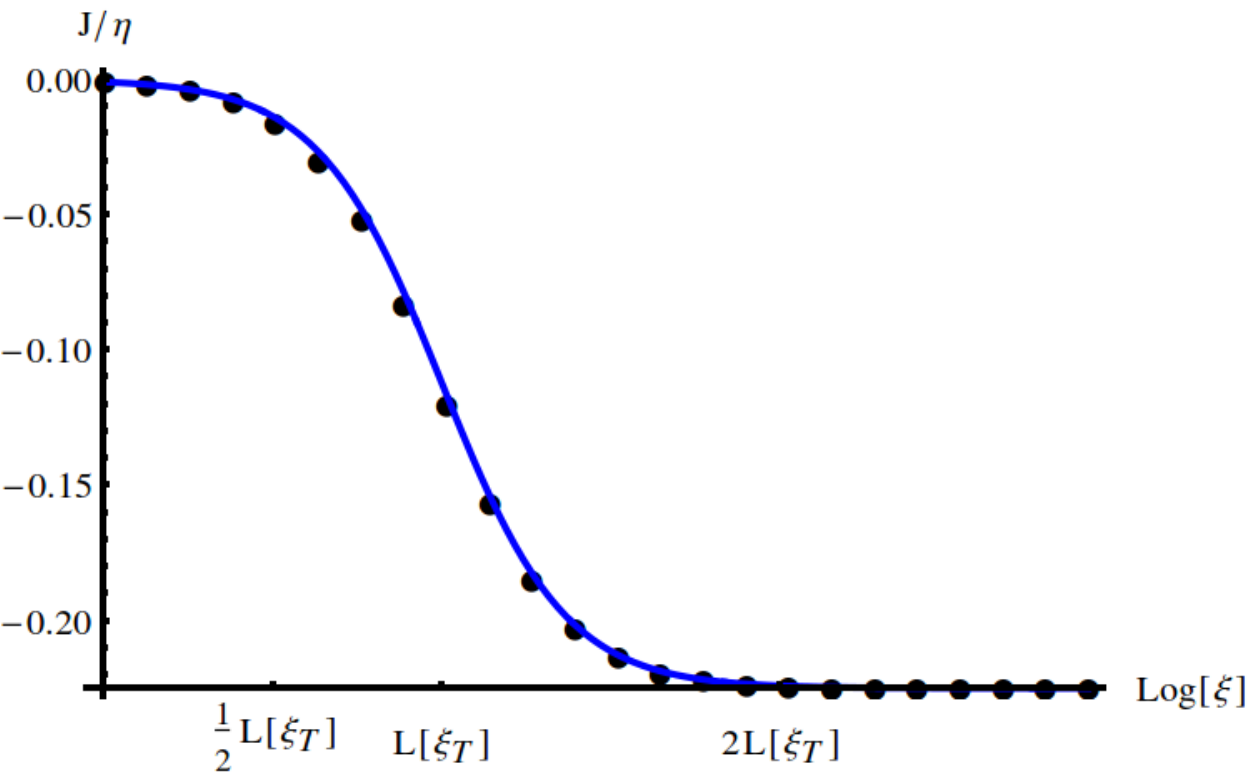}\protect\protect\protect\protect\protect\protect\caption{Comparison of the analytic approximation to $J_{\mathrm{fluct}}^{C}$
given by Eq. (\ref{J_CDW_approx}) (solid blue line) with the numerical
evaluation of the corresponding integral in Eq. (\ref{J_evaluated})
(black dots). For convenience, we defined $L(\xi_{T})\equiv\log(\xi_{T})$.}

\label{FigCDWNumericsVSAssymptotics} 
\end{figure}

\begin{equation}
J_{\mathrm{fluct}}^{C}\approx\left[\frac{2\braket{\vec{v}_{F}^{2}(\phi)}_{\phi}}{\pi}\right]\frac{C_{\mathrm{C}}\eta_{C}}{\left(\frac{\xi_{T}}{\bar{\xi}_{\mathrm{C}}}\right)^{2}+1}\label{J_CDW_approx}
\end{equation}
with the negative geometrical pre-factor: 
\begin{equation}
C_{\mathrm{C}}=-\left(\frac{N_{\phi_{1}}}{\nu_{F}}\right)^{2}\frac{v_{x}^{2}(\phi_{1}^{x})}{2\braket{\vec{v}_{F}^{2}(\phi)}_{\phi}\left|\pd{p_{x}}{\phi_{1}}\pd{p_{y}}{\phi_{1}}\right|}
\end{equation}

In order to estimate the precision of the asymptotic result obtained
above we compared it with the corresponding numerical evaluation of
$J_{\mathrm{fluct}}^{C}$ in Fig.~\ref{FigCDWNumericsVSAssymptotics}
using the tight-binding dispersion of Ref. \cite{Das12} with $\mu=0.1$.
We checked that the agreement is robust for changes in the chemical
potential and also for other tight-binding dispersions, such as that
used in Ref. \cite{Sachdev_LaPlaca}.

\subsubsection{SDW fluctuations}

The contribution from the SDW fluctuations are more involved due to
the higher symmetry of the fluctuations. The hot spots, connected
by the ordering vector $\mathbf{Q}_{S}=\left(\pi,\pi\right)$, correspond
to the angles $\phi_{1}$, $\phi_{2}=\frac{3\pi}{2}-\phi_{1}$. Expansion
around these angles gives:

\begin{align}
q_{x}-Q_{x} & \approx\pd{p_{x}}{\phi_{1}}\delta\phi_{1}-\pd{p_{y}}{\phi_{1}}\delta\phi_{2}+\frac{1}{2}\pdn{p_{x}}{\phi_{1}}{2}\delta\phi_{1}^{2}+\frac{1}{2}\pdn{p_{y}}{\phi_{1}}{2}\delta\phi_{2}^{2}\\
q_{y}-Q_{y} & \approx\pd{p_{y}}{\phi_{1}}\delta\phi_{1}-\pd{p_{x}}{\phi_{1}}\delta\phi_{2}+\frac{1}{2}\pdn{p_{y}}{\phi_{1}}{2}\delta\phi_{1}^{2}+\frac{1}{2}\pdn{p_{x}}{\phi_{1}}{2}\delta\phi_{2}^{2}
\end{align}

Unlike the CDW case, an expansion only in the denominator of Eq. (\ref{J_evaluated})
is however not enough, because the anisotropy comes from the momentum-dependent
part of the susceptibility. Therefore, we expand also the angle-dependent
DOS:

\begin{equation}
N_{\phi}N_{\phi'}\approx N_{\phi_{1}}^{2}\left[1+\left(\delta\phi_{1}-\delta\phi_{2}\right)\pd{}{\phi}\log(N_{\phi})\right]=N_{\phi_{1}}^{2}\left[1+\mu_{\phi_{1}}\left(\delta\phi_{1}-\delta\phi_{2}\right)\right]
\end{equation}
as well as the velocity combination:

\begin{equation}
\left(v_{x}(\phi_{1}+\delta\phi_{1})-v_{x}(\phi_{2}+\delta\phi_{2})\right)^{2}-\left(v_{y}(\phi_{1}+\delta\phi_{1})-v_{y}(\phi_{2}+\delta\phi_{2})\right)^{2}\approx\left(\delta\phi_{1}+\delta\phi_{2}\right)\left[A_{1}+A_{2}\left(\delta\phi_{1}-\delta\phi_{2}\right)\right]
\end{equation}
with 
\begin{align}
A_{1} & =2\left(v_{x}(\phi_{1})+v_{y}(\phi_{1})\right)\left(\pd{v_{x}}{\phi_{1}}-\pd{v_{y}}{\phi_{1}}\right)\\
A_{2} & =\left[\left(\pd{v_{x}}{\phi_{1}}\right)^{2}-\left(\pd{v_{y}}{\phi_{1}}\right)^{2}+\left(v_{x}(\phi_{1})+v_{y}(\phi_{1})\right)\left(\pdn{v_{x}}{\phi_{1}}{2}-\pdn{v_{y}}{\phi_{1}}{2}\right)\right].
\end{align}

It is straightforward to recognize that the transformation $(\delta\phi_{1},\delta\phi_{2})\leftrightarrow-(\delta\phi_{2},\delta\phi_{1})$
is equivalent to $(q_{y}-Q_{y})\leftrightarrow(q_{x}-Q_{x})$. Since
this transformation does not alter the measure and merely changes
the global sign of the velocity part we split $\tilde{\omega}_{q}$
into odd ($L$) and even ($K$) parts with respect to the transformation
above: 
\begin{equation}
\tilde{\omega}_{q}=\xi_{\mathrm{S}}^{-2}+K_{1}+K_{2}+\eta_{\mathrm{S}}(L_{1}+L_{2})
\end{equation}
with the leading order ($K_{1}$, $L_{1}$) and next-to-leading order
($K_{2}$, $L_{2}$) contributions:

\begin{align}
K_{1} & =\frac{1}{2}\left(\pd{p_{x}}{\phi_{1}}+\pd{p_{y}}{\phi_{1}}\right)^{2}\left(\delta\phi_{1}-\delta\phi_{2}\right)^{2}+\frac{1}{2}\left(\pd{p_{x}}{\phi_{1}}-\pd{p_{y}}{\phi_{1}}\right)^{2}\left(\delta\phi_{1}+\delta\phi_{2}\right)^{2}\\
K_{2} & =\left(\pd{p_{x}}{\phi_{1}}\pdn{p_{x}}{\phi_{1}}{2}+\pd{p_{y}}{\phi_{1}}\pdn{p_{y}}{\phi_{1}}{2}\right)\left(\delta\phi_{1}^{3}-\delta\phi_{2}^{3}\right)-\left(\pd{p_{x}}{\phi_{1}}\pdn{p_{y}}{\phi_{1}}{2}+\pd{p_{y}}{\phi_{1}}\pdn{p_{x}}{\phi_{1}}{2}\right)\delta\phi_{1}\delta\phi_{2}\left(\delta\phi_{1}-\delta\phi_{2}\right)\\
L_{1} & =\left[\left(\pd{p_{x}}{\phi_{1}}\right)^{2}-\left(\pd{p_{y}}{\phi_{1}}\right)^{2}\right]\left(\delta\phi_{1}^{2}-\delta\phi_{2}^{2}\right)\\
L_{2} & =\left(\pd{p_{x}}{\phi_{1}}\pdn{p_{x}}{\phi_{1}}{2}-\pd{p_{y}}{\phi_{1}}\pdn{p_{y}}{\phi_{1}}{2}\right)\left(\delta\phi_{1}^{3}+\delta\phi_{2}^{3}\right)+\left(\pd{p_{x}}{\phi_{1}}\pdn{p_{y}}{\phi_{1}}{2}-\pd{p_{y}}{\phi_{1}}\pdn{p_{x}}{\phi_{1}}{2}\right)\delta\phi_{1}\delta\phi_{2}\left(\delta\phi_{1}+\delta\phi_{2}\right)
\end{align}

Similarly to the CDW case, we define $r=\delta\phi_{1}+\delta\phi_{2}$
and $s=\delta\phi_{1}-\delta\phi_{2}$. Expanding to the lowest order
of $\eta_{\mathrm{S}}$ we find (for notation simplicity $L_{i}=L_{i}(r,s)$
and $K_{i}=K_{i}(r,s)$):

\begin{align}
J_{\mathrm{fluct}}^{S}\approx & -\eta_{S}\left(\frac{N_{\phi_{1}}}{\nu_{F}}\right)^{2}\frac{2\xi_{T}^{-2}}{\pi^{2}}\int\mathrm{d}r\int\mathrm{d}s\frac{\left(L_{1}+L_{2}\right)\left(\xi_{S}^{-2}+\frac{1}{2}\xi_{T}^{-2}+K_{1}+K_{2}\right)\left(1+\mu s\right)r\left(A_{1}+A_{2}s\right)}{\left[\left(\xi_{S}^{-2}+K_{1}+K_{2}\right)^{2}\right]\left[\left(\xi_{S}^{-2}+\xi_{T}^{-2}+K_{1}+K_{2}\right)^{2}\right]}\\
J_{\mathrm{fluct}}^{S}\approx & \left[\frac{2\braket{\vec{v}_{F}^{2}(\phi)}_{\phi}}{\pi}\right]\eta_{S}C_{S}\xi_{T}^{-2}\left\{ \log\left(\xi_{\mathrm{S}}^{2}\right)+\left[1-\left(1+\frac{\xi_{T}^{2}}{\xi_{\mathrm{S}}^{2}}\right)\log\left(1+\frac{\xi_{\mathrm{S}}^{2}}{\xi_{T}^{2}}\right)\right]\right\} \label{J_SDW_approx}
\end{align}
with the geometrical pre-factor: 
\begin{equation}
C_{S}=-\left(\frac{N_{\phi_{1}}}{\nu_{F}}\right)^{2}\frac{\left[\left(\pd{p_{x}}{\phi_{1}}\right)^{2}-\left(\pd{p_{y}}{\phi_{1}}\right)^{2}\right]2\left(\mu A_{1}+A_{2}\right)-A_{1}\left[3\left(\pd{p_{x}}{\phi_{1}}\pdn{p_{x}}{\phi_{1}}{2}-\pd{p_{y}}{\phi_{1}}\pdn{p_{y}}{\phi_{1}}{2}\right)-\left(\pd{p_{x}}{\phi_{1}}\pdn{p_{y}}{\phi_{1}}{2}-\pd{p_{y}}{\phi_{1}}\pdn{p_{x}}{\phi_{1}}{2}\right)\right]}{2\braket{\vec{v}_{F}^{2}(\phi)}_{\phi}\left|\left(\pd{p_{x}}{\phi_{1}}\right)^{2}-\left(\pd{p_{y}}{\phi_{1}}\right)^{2}\right|^{3}}
\end{equation}

Because the SDW fluctuations are only anisotropic in their momentum
dependence, the geometrical pre-factor is no longer determined only
by the hot-spots Fermi velocity, but also by their derivatives and
the curvature of the Fermi surface near the hot spots. As a result,
it is a priori not clear what the sign of $C_{S}$ is. We evaluated
$C_{S}$ numerically using the tight binding models of Refs.~\cite{Das12,Sachdev_LaPlaca}
and found $C_{\mathrm{S}}=-0.25$~ for Ref. \cite{Sachdev_LaPlaca}
(with chemical potential $\mu=-1.3$) and $C_{\mathrm{S}}=-0.067$
for Ref. \cite{Das12} (with chemical potential $\mu=0.1$). We also
found that the negative character of $C_{S}$ is robust against small
changes in the chemical potential. Thus, we observe in general a tendency
towards $C_{S}<0$. The only situation in which we were able to find
$C_{S}>0$ was for Fermi surface configurations that tend to become
open around $\left(\pi,\pi\right)$ (closed near $\left(0,0\right)$),
which are not relevant for hole-underdoped compounds.

\begin{figure}
\includegraphics[width=0.35\columnwidth]{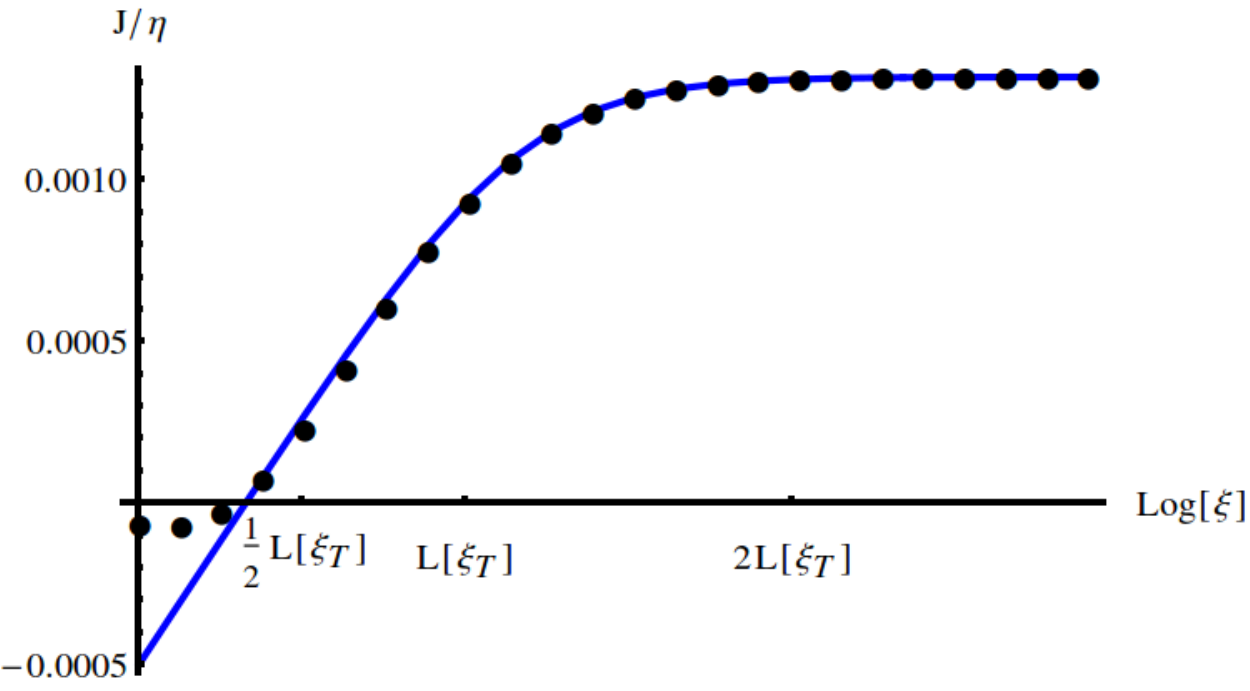}\protect\protect\protect\protect\protect\protect\caption{Comparison of the analytic approximation to $J_{\mathrm{fluct}}^{C}$
given by Eq. (\ref{J_SDW_approx}) (solid blue line) with the numerical
evaluation of the corresponding integral in Eq. (\ref{J_evaluated})
(black dots). For convenience, we defined $L(\xi_{T})\equiv\log(\xi_{T})$.}

\label{FigSDWNumericsVSAssymptotics} 
\end{figure}

Note that, in Eq. (\ref{J_SDW_approx}), $J_{\mathrm{fluct}}^{S}$
saturates in the regime $\xi_{S}\gg\xi_{T}$. Therefore, it is in
principle possible that other contributions to $J_{\mathrm{fluct}}^{S}$
not related to the hot spots are comparable to those arising from
the hot spots physics. To check this, we computed numerically Eq.
(\ref{J_evaluated}) for the tight-binding models of Refs.~\cite{Das12,Sachdev_LaPlaca}
for a range of chemical potential values. In Fig.~\ref{FigSDWNumericsVSAssymptotics}
we show the specific case of the tight-binding parameters of Ref.
\cite{Das12} with $\mu=0.1$, comparing it with the analytical expression
given by Eq. (\ref{J_SDW_approx}). To account for the contribution
that does not arise from the hot spots physics, we added a constant
shift. Clearly, this additional contribution is smaller than that
from the hot-spots for $\xi_{S}\gg1$. We found a similar behavior
when using the tight-binding parameters of Refs.~\cite{Das12,Sachdev_LaPlaca}
for a variety of chemical potential values, demonstrating that Eq.
(\ref{J_SDW_approx}) captures the behavior of the resistivity anisotropy
in the regime where $\xi_{S}$ is not too small.\end{widetext}

\end{document}